\documentclass[12pt]{iopart}

\usepackage{graphicx,setspace}
\usepackage{dsfont}
\usepackage{iopams}
\usepackage{mathrsfs}
\usepackage{psfrag}
\usepackage{cases}

\begin{document}
\title{Damping the zero-point energy of a harmonic oscillator}

\author{T G Philbin and S A R Horsley}
\address{Physics and Astronomy Department, University of Exeter,
Stocker Road, Exeter EX4 4QL, UK.}
\eads{\mailto{t.g.philbin@exeter.ac.uk}, \mailto{s.horsley@exeter.ac.uk}}

\begin{abstract}
The physics of quantum electromagnetism in an absorbing medium is that of a field of damped harmonic oscillators.  Yet until recently the damped harmonic oscillator was not treated with the same kind of formalism used to describe quantum electrodynamics in a arbitrary medium.  Here we use the techniques of macroscopic QED, based on the Huttner--Barnett reservoir, to describe the quantum mechanics of a damped oscillator. We calculate the thermal and zero-point energy of the oscillator for a range of damping values from zero to infinity. While both the thermal and zero-point energies decrease with damping, the energy stored in the oscillator at fixed temperature \emph{increases} with damping, an effect that may be experimentally observable. As the results follow from canonical quantization, the uncertainty principle is valid for all damping levels.
\end{abstract}
\pacs{03.65.-w, 03.65.Yz, 03.70.+k}

\maketitle

\section{Introduction}
The damped harmonic oscillator has a central place in physics, due to the prevalance of dissipation and linear response in our description of the physical world.  Yet while the classical treatment of the damped oscillator is elementary and well understood, the opposite is true in quantum physics. The dissipation of energy leads to difficulties in applying the standard quantization rules to the damped oscillator~\cite{dek81,um02,wei08,bat31,gra84,bla02,lat05,bal11,maj12,mag59,fey63,cal83,tat87,smi90,han05,han08,ing09a,ing09b,dat10,ing12,phi12}.  Moreover, current experimental work is probing the degree to which quantum mechanics may describe the macroscopic world~\cite{oco10,teu11,cha11,asp10,poo12}.  In these experiments the regime is often one where linear response is applicable and the degree of dissipation is of interest.  It is not yet clear to what extent existing approaches to quantum damped systems accurately describe the results~\cite{gro13}.  Here we explore the consequences of a recent, and particularly general treatment of the damped harmonic oscillator~\cite{phi12}, where a measured value of the dissipative response of the system as a function of frequency may be input as a parameter directly into the Hamiltonian, something which is not possible in previous approaches.  The thought is that, in analogy to electromagnetism, where the susceptibilities of a medium are routinely measured and then used as parameters within the quantum theory~\cite{sch09,phi10}, one might extend this approach to describe the aforementioned experiments.  Through choosing a model susceptibility where the system is exactly solvable, we find the interesting result that the ground state energy of a damped oscillator---as calculated using the Hamiltonian of mean force---decreases as the damping is increased, and propose a means whereby this phenomenon might be measured.

The technical difficulties in the canonical quantization of the damped harmonic oscillator can be overcome through the inclusion of reservoir degrees of freedom that take up the dissipated energy. If the reservoir has a finite, or countably infinite, number of degrees of freedom, then a delicate limiting procedure must be employed to try to capture genuine damping behaviour~\cite{wei08,mag59,fey63,cal83,tat87,smi90,han05,han08,ing09a,ing09b,dat10,ing12}. This limiting procedure is not usually performed in detail, but the subtleties involved are lucidly demonstrated by Tatarskii~\cite{tat87}. The limiting procedure amounts to a transition from a finite to an uncountably infinite number of dynamical degrees of freedom in the reservoir. The fact that this limit is imposed after the field equations have already been solved under the assumption of a \emph{finite} reservoir obscures any connection with the original Hamiltonian. Indeed, it has been shown~\cite{tat87} that the limiting procedure must be separately constructed for each specific damping behaviour if it is to be rigorously performed, something that is generally dispensed with in practice. Rather than follow this well-trodden path, here we apply the technique used to quantize electromagnetism in absorbing media~\cite{hut92,phi10}, where a reservoir with an uncountably infinite number of degrees of freedom is used from the outset. This method was extended to the damped oscillator in~\cite{phi12} and has the advantage that the imaginary part of the susceptibility that governs the dynamics of the oscillator appears explicitly as a parameter in the Lagrangian or Hamiltonian. There is no limiting procedure to be performed and moreover the quantum dynamics of the damped oscillator is then placed on the same footing as macroscopic quantum electrodynamics. In what follows we use this approach to explore the behaviour of the thermal and zero-point energy of the quantum damped oscillator. We apply the previously derived results~\cite{phi12} for a general damped oscillator in thermal equilibrium to a simple model susceptibility that allows exact analytic solution for all quantities of interest. This example is used as a guide to the type of effects that may be measurable with current macroscopic quantum oscillators. We find that the zero-point energy of the oscillator is less than $\hbar\omega_0/2$, where $\omega_0$ is the free oscillation frequency in the absence of damping. The energy removed in cooling the oscillator from temperature $T>0$ to its quantum ground state is found to increase with damping, which offers one possibility of experimentally demonstrating the effect derived here.  It is also plausible that the damping might be engineered, as we discuss below. 

\section{Thermal equilibrium and a choice of susceptibility}
The form of the Lagrangian for a single damped degree of freedom, \(q(t)\) (unit mass and free oscillation frequency \(\omega_{0}\)) was previously given as~\cite{phi12}
\begin{equation}
	L=\frac{1}{2}\left(\dot{q}^{2}-\omega_{0}^{2}q^{2}\right)+q\int_{0}^{\infty}d\omega\,\alpha(\omega)X_{\omega}
	+\frac{1}{2}\int_{0}^{\infty}d\omega\left(\dot{X}_{\omega}^{2}-\omega^{2}X_{\omega}^{2}\right),     \label{lagrangian}
\end{equation}
where the \(X_{\omega}\) constitute the reservoir and are labelled by a continuum `index' \(\omega\).  The coupling function $\alpha(\omega)$ between reservoir and oscillator is related to the imaginary part of the susceptibility by 
\begin{equation} \label{algen}
\alpha(\omega)=\omega_{0}\sqrt{\frac{2\omega\mathrm{Im}[\chi(\omega)]}{\pi}},
\end{equation} 
where $\chi(\omega)$ is the linear susceptibility that quantifies the effect of the environment on the motion of the oscillator (the factor of \(\omega_{0}\) in \(\alpha(\omega)\) is present so that the susceptibility is dimensionless). The susceptibility $\chi(\omega)$ obeys the Kramer-Kronig relations~\cite{phi12}, so the imaginary part of $\chi(\omega)$ that appears in the Lagrangian (\ref{lagrangian}) determines the real part. In fact the Kramers-Kronig relations give the full susceptibility in terms of the coupling function $\alpha(\omega)$ as~\cite{phi12} 
\begin{equation}  \label{chireim}
	\omega_0^2\,\chi(\omega)=\mathrm{P}\int_{0}^{\infty} d \xi\frac{\alpha^2(\xi)}{\xi^2-{\omega}^2}+\frac{i\pi \alpha^2(\omega)}{2\omega}.
\end{equation}
The resulting theory is more transparent than Caldeira--Leggett type approaches, in that the reservoir stipulated in the Lagrangian is sufficient for the task and does not have to be modified later. In fact, the approach in (\ref{lagrangian}) is essentially identical to that used in quantum electromagnetism within an arbitrary absorbing dielectric medium, which can of course be viewed as a field theory of damped oscillators. In particular, the dynamics of a general damped harmonic oscillator is governed by an arbitrary susceptibility that obeys the Kramers-Kronig relations~\cite{phi12}, just as for the dynamics of light in a material medium. Our point of view is that canonical quantization of macroscopic electromagnetism for arbitrary dielectrics~\cite{phi10,hor11} and canonical quantization of a general damped harmonic oscillator~\cite{phi12} are both most effectively carried out using the powerful reservoir formalism originally introduced by Huttner and Barnett~\cite{hut92}. 

From (\ref{lagrangian}) we can derive a Hamiltonian and apply the standard quantization rules.  The position operator for the oscillator is given by \(\hat{q}\), the canonical momentum by $\hat{\Pi}_q(t)$, and  $[\hat{q}(t),\hat{\Pi}_q(t)]=i\hbar$ is always satisfied~\cite{phi12}. In thermal equilibrium the expectation values of the squares of the position and momentum operators were previously shown to be~\cite{phi12}
\begin{eqnarray}  
	\left\langle \hat{q}^2(t)\right\rangle=&\frac{\hbar}{\pi}\int_0^\infty d\omega\coth\left(\frac{\hbar\omega}{2k_BT}\right) \mathrm{Im}G(\omega),  \label{qqtime} \\
	\left\langle \hat{\Pi}^2_q(t)\right\rangle=&\frac{\hbar}{\pi}\int_0^\infty  d  \omega \,\omega^2\coth\left(\frac{\hbar\omega}{2k_BT}\right) \mathrm{Im}G(\omega) , \label{PqPqtime}
\end{eqnarray}
where $G(\omega)$ is the Green function for the motion of the oscillator, containing the susceptibility $\chi(\omega)$:
\begin{equation}  \label{Gchi}
	G(\omega)=\frac{-1}{\omega^2-\omega_0^2\left[1-\chi(\omega)\right]}.
\end{equation}
When the damping is zero the  susceptibility $\chi(\omega)$ vanishes and the Green function (\ref{Gchi}) is that of a free oscillator of frequency $\omega_0$.
The results (\ref{qqtime}) and (\ref{PqPqtime}) could have been anticipated from the fluctuation--dissipation theorem~\cite{vol5}, but here they are derived~\cite{phi12} from the Lagrangian (\ref{lagrangian}), illustrating the consistency of our approach with known results from statistical physics. 

The energy of the oscillator in thermal equilibrium cannot be anticipated from the fluctuation--dissipation theorem, but must be obtained from the total thermal energy of the coupled oscillator/reservoir system.  This was calculated in~\cite{phi12} by subtracting, from the total thermal energy, the thermal energy of the reservoir in the absence of any coupling to the oscillator, giving the result
\begin{equation}   
\fl
	\langle\hat{H}\rangle_q= \frac{\hbar}{2\pi}\int_{0}^\infty d\omega  \coth\left(\frac{\hbar\omega}{2k_BT}\right) 
 	 \mathrm{Im}\left\{\left[\omega_0^2\left(\omega\frac{d\chi(\omega)}{d\omega}-\chi(\omega)+1\right)+\omega^2\right]G(\omega)\right\}. \label{Hq}
\end{equation}
A similar procedure in the electromagnetic case gives the Casimir (zero-point plus thermal) stress-energy of the electromagnetic field in a material~\cite{phi10}. In the Appendix we show that the prescription giving (\ref{Hq}) is equivalent to the thermal average of the Hamiltonian of mean force~\cite{cam09}, consistent with earlier work on the thermodynamics of strongly coupled systems~\cite{for85}.  In light of this, the results of~\cite{phi10} show that the Casimir energy density is the thermal average of the Hamiltonian of mean force for the electromagnetic field in a macroscopic medium, a connection that was not recognised in~\cite{phi10} and which does not appear to be widely appreciated.

In accordance with our viewpoint that damping ought to be generally treated in the same way as electromagnetic dissipation, the susceptibility $\chi(\omega)$ of a real damped oscillator should be measured rather than postulated.  Recall that the electromagnetic susceptibility of an individual material sample must be measured, and will vary even for samples of the same material. The position and momentum correlation functions~\cite{phi12} of the damped oscillator in thermal equilibrium provide one method of experimentally extracting the quantities $\chi(\omega)$ and $\omega_0$ that appear in (\ref{Gchi}).

In the absence of tabulated values, it is instructive to consider simple formulae for the susceptibility of a damped oscillator.  In~\cite{phi12} the example of damping proportional to velocity was treated in detail, but this gives some problems with the zero-damping limit at $T>0$ if the corresponding susceptibility $\chi(\omega)$ is taken to hold strictly at all frequencies up to infinity. In addition, the case of damping proportional to velocity would not be expected to be experimentally relevant~\cite{phi12}. We therefore consider another example, chosen to be simple enough to allow exact analytical solution while being well-behaved in the limit of zero damping of the oscillator. Our particular susceptibility takes the form
\begin{equation}  \label{chi}
	\chi(\omega)=\frac{2\gamma_2(\gamma_1^2+\omega_0^2)}{\omega_0^2(\gamma_1+2\gamma_2-i\omega)},
\end{equation}
where $\gamma_1$ and $\gamma_2$ are positive real constants. Being analytic in the upper-half complex-frequency plane, the real and imaginary parts of (\ref{chi}) exhibit Kramer-Kronig relations and from (\ref{algen}) and (\ref{chireim}) we find that the coupling function $\alpha(\omega)$ in (\ref{lagrangian}) corresponding to the susceptibility (\ref{chi}) is
\begin{equation}  \label{al}
	\alpha(\omega)=\sqrt{\frac{4\gamma_2\omega^2(\gamma_1^2+\omega_0^2)}{\pi[(\gamma_1+2\gamma_2)^2+\omega^2]}}.
\end{equation}
It is important to note that the thermal results (\ref{qqtime}), (\ref{PqPqtime}) and (\ref{Hq}) are only valid for cases where the total Hamiltonian can be diagonalized into normal modes, which is not always possible~\cite{phi12}.  A sufficient condition for the Hamiltonian to be diagonalizable was found in~\cite{phi12} to be
\begin{equation} \label{ineqgen}
 	\omega_0^2>\int_0^\infty d \xi  \frac{\alpha^2(\xi)}{\xi^2},
\end{equation}
a restriction that would not be transparent from the fluctuation--dissipation theorem.  For the coupling function (\ref{al}), the condition (\ref{ineqgen}) yields
\begin{equation} \label{ineq}
	\omega_0^2>2\gamma_1\gamma_2,
\end{equation}
which we assume to hold throughout. When $\gamma_2\to0$, $\gamma_1$ remaining fixed, the susceptibility (\ref{chi}) vanishes and the case of an undamped oscillator is recovered.
\par
The Green function (\ref{Gchi}) for the susceptibility (\ref{chi}) has poles 
\begin{eqnarray}
\omega=-i\gamma_1, \quad \omega=-i\gamma_2\pm\omega_{1},  \label{Gpoles} \\
\omega_1=\sqrt{ \omega_0^2-\gamma_2(2\gamma_1+\gamma_2)},  \label{omega1}
\end{eqnarray}
which correspond to the eigenfrequencies of the oscillator coupled to the reservoir.  The constants $\gamma_1$ and $\gamma_2$ thus serve as damping constants of the oscillator, while $\omega_1$ is a modified oscillation frequency when it is real. Note that (\ref{ineq}) implies that the poles (\ref{Gpoles}) are all in the lower-half complex-frequency plane so the Green function has retarded boundary conditions. Note also that the over-damped case with imaginary $\omega_1$ can occur while still satisfying (\ref{ineq}).

\section{Thermal and zero-point results}
\begin{figure}
\begin{center}
\includegraphics[width=9cm]{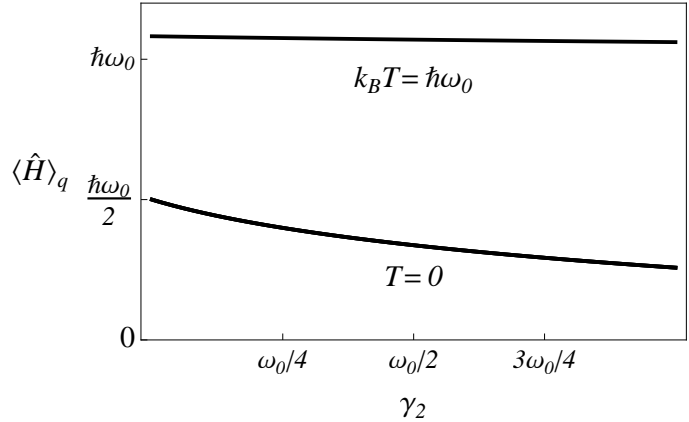}
\caption{Plots of the energy of the harmonic oscillator versus damping $\gamma_2$, with $\omega_0=10^{10}\,\mathrm{s}^{-1}$ and $\gamma_1=\omega_0/4$, for $T=0$ and $T=\hbar\omega_0/k_B$. The zero-point energy ($T=0$) is damped below the free-oscillator value $\hbar\omega_0/2$. The energy for $T>0$ is also damped below the free-oscillator value, though this damping is not very apparent except for very low $T$. The energy that can be extracted from the oscillator at $T>0$ (i.e.\ the $T>0$ energy minus the zero-point energy) increases with damping. } \label{fig:zpe}
\end{center} 
\end{figure}

The thermal expectation values (\ref{qqtime}), (\ref{PqPqtime}) and (\ref{Hq}) can all be evaluated exactly for the susceptibility (\ref{chi}). The integrands in each case are even functions of $\omega$ for $T>0$ and so can be rewritten with lower integration limit of $-\infty$; the integrals are then evaluated by closing the integration contour in the upper (or lower) half-plane. The infinite sum over the residues of the poles of the hyperbolic cotangent function can be evaluated exactly, but the resulting expressions are rather lengthy and we do not give them here. In the limit $\gamma_2\to0$ we find the expectation values (\ref{qqtime}), (\ref{PqPqtime}) and (\ref{Hq}) reduce to the free-oscillator values $(\hbar/2\omega_0)\mathrm{coth}(\hbar\omega_0/2k_BT)$, $(\hbar\omega_0/2)\mathrm{coth}(\hbar\omega_0/2k_BT)$ and $(\hbar\omega_0/2)\mathrm{coth}(\hbar\omega_0/2k_BT)$, respectively (by the virial theorem, the momentum-squared expectation value is equal to the thermal energy for a free oscillator with unit mass). The thermal energy as a function of the damping $\gamma_2$ is plotted for temperature $T=\hbar\omega_0/k_B$ in Figure~\ref{fig:zpe}.

The zero-point ($T=0$) values of (\ref{qqtime}), (\ref{PqPqtime}) and (\ref{Hq}) take a simpler form than the thermal results. They are best evaluated separately rather as the $T\to0$ limit of the thermal case, and we find them to be
\begin{eqnarray}  
\fl
\left\langle \hat{q}^2(t)\right\rangle=  \frac{\hbar}{\pi\omega_1(\omega_0^2+\gamma_1^2-4\gamma_1\gamma_2)}    \nonumber   \\  
\fl
 \qquad\qquad \times\left[(\omega_1^2+\gamma_1^2-\gamma_2^2)\mathrm{arctan}\left(\frac{\omega_1}{\gamma_2}\right) + \gamma_2 \omega_1\ln\left(\frac{\omega_0^2-2 \gamma_1\gamma_2}{\gamma_1^2}\right)      \right] ,   \label{qqzp} \\
 \fl
\left\langle \hat{\Pi}^2_q(t)\right\rangle=   \frac{\hbar}{\pi\omega_1(\omega_0^2+\gamma_1^2-4\gamma_1\gamma_2)}    \nonumber   \\
\fl
 \qquad\qquad\   \times \left\{\left[(\omega_1^2+\gamma_2^2)^2+\gamma_1^2(\omega_1^2-\gamma_2^2)\right]\mathrm{arctan}\left(\frac{\omega_1}{\gamma_2}\right) - \gamma_1^2\gamma_2 \omega_1\ln\left(\frac{\omega_0^2-2 \gamma_1\gamma_2}{\gamma_1^2}\right)      \right\},   \label{PqPqzp}   \\
\fl
\langle\hat{H}\rangle_q=    \frac{\hbar}{2\pi}\left\{ 2\omega_1\mathrm{arctan}\left(\frac{\omega_1}{\gamma_2}\right)+ \gamma_1\ln\left(1+2\frac{\gamma_2}{\gamma_1}\right)+\gamma_2\ln\left[\frac{(\gamma_1+2\gamma_2)^2}{\omega_0^2-2 \gamma_1\gamma_2}\right]   \right\} .  \label{Hqzp}
\end{eqnarray}
Recall that these expressions presuppose the inequality (\ref{ineq}) and note that they are real in the over-damped case where $\omega_1$ is imaginary.  We now consider the dependence of (\ref{qqzp})--(\ref{Hqzp}) on the parameters within the susceptibility.

\begin{figure}
\begin{center}
\includegraphics[width=9cm]{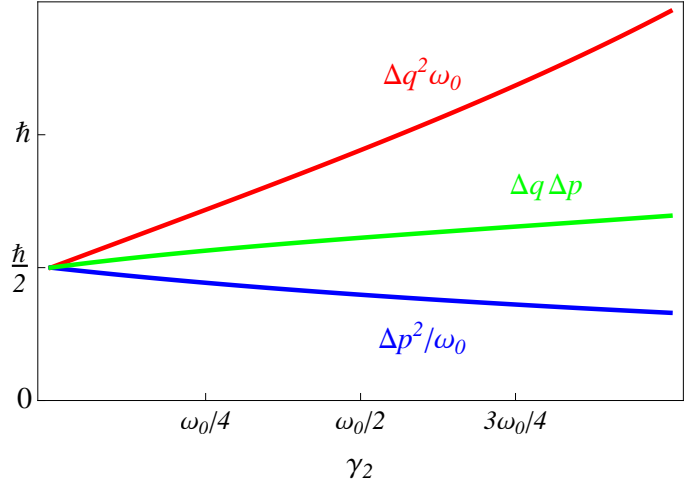}
\caption{Plots of the zero-point position uncertainty $\Delta q$ (square root of (\ref{qqzp})) and momentum uncertainty $\Delta p$ (square root of (\ref{PqPqzp})) versus damping $\gamma_2$ with $\omega_0=10^{10}\,\mathrm{s}^{-1}$ and $\gamma_1=\omega_0/4$. The squares of the uncertainties are scaled with an appropriate power of $\omega_0$ to have the same units as the product $\Delta q\,\Delta p$. The uncertainty relation is satisfied for all parameters obeying (\ref{ineq}). } \label{fig:up}
\end{center} 
\end{figure}

Figure~\ref{fig:up} shows the uncertainties in position and momentum,  \(\Delta q\) and \(\Delta p\)  (the square roots of (\ref{qqzp}) and (\ref{PqPqzp})), as functions of $\gamma_2$ for $\omega_0=10^{10}\,\mathrm{s}^{-1}$ and $\gamma_1=\omega_0/4$. As damping ($\gamma_2$) increases the product $\Delta q\,\Delta p$ increases from the minimum allowed value $\hbar/2$. The position uncertainty $\Delta q$ increases with damping while $\Delta p$ decreases.  The observed adherence to the position/momentum uncertainty relation is unsurprising, given that the results are based on canonical quantization, but violations of the uncertainty principle occur in other approaches to the damped oscillator~\cite{dek81,um02}.

The zero-point energy (\ref{Hqzp}) of the oscillator is plotted in Figure~\ref{fig:zpe} as a function of $\gamma_2$ for the same values of \(\omega_{0}\) and \(\gamma_{1}\) used in Figure~\ref{fig:up}; the energy at temperature $T=\hbar\omega_0/k_B$ is also plotted. As $\gamma_2$ increases the zero-point energy is damped below the free-oscillator value $\hbar\omega_0/2$.  In our example, the oscillation frequency for $\gamma_2>0$ is $\omega_1$, given by (\ref{omega1}), provided  $\omega_1$ is real.  We emphasise, as is clear from (\ref{Hqzp}), that the zero-point energy of the damped oscillator is not $\hbar\omega_1/2$.  In fact the range of $\gamma_2$ in Figure~\ref{fig:zpe} passes through $\omega_1=0$ and into the over-damped case where $\omega_1$ is imaginary.  The rather complicated zero-point energy (\ref{Hqzp}) thus cannot be simply related to the oscillation behaviour, given by (\ref{Gpoles}) and (\ref{omega1}), although both are affected by the damping. The energy at $T>0$ is also damped below the free-oscillator value but the difference between the $T>0$ energy and the zero-point energy increases with damping.  This shows that the energy stored in the oscillator at fixed $T$ increases with damping, so that an increasing quantity of energy must be removed to bring the oscillator to its ground state. 

We can also consider the limit of infinite damping, which occurs when $\gamma_2\to\infty$. The condition (\ref{ineq}) then requires $\gamma_1\to0$, which we can satisfy by setting $\gamma_1= \omega_0^2/(4\gamma_2)$ for example. With this value for $\gamma_1$, the zero-point energy (\ref{Hqzp}) goes to zero for infinite damping $\gamma_2\to\infty$, with a leading term of 
\begin{equation}
\langle\hat{H}\rangle_q\sim\frac{\hbar\omega_0^2}{4\pi\gamma_2}\left[1+2\ln\left(\frac{2^{3/2}\gamma_2}{\omega_0}\right)\right].
\end{equation}
The asymptotic approach of the zero-point energy to zero as $\gamma_2\to\infty$ is thus very slow. We must note also that the position uncertainty diverges as $\Delta q\sim2\sqrt{\hbar\gamma_2/\pi}/\omega_0$ in this infinite-damping limit (with $\gamma_1= \omega_0^2/(4\gamma_2)$). A large displacement of the oscillator can be expected to lead to nonlinear behaviour, so our assumption of a linear oscillator is not realistic for extremely large damping with susceptibility (\ref{chi}).  

As noted at the outset, the susceptibility is a quantity that must be measured, and in addition the ``free-oscillation" frequency $\omega_0$ is a parameter that must also be fitted to experimental data~\cite{phi12}. The theoretical ideal is an oscillator whose damping can be tuned from zero to a desired level, but this is a heavy demand in practice. A more realistic scenario is a set of macroscopic oscillators prepared with slightly different material geometries so that the damping varies slowly across the set. In the absence of data for the susceptibilities of such a set of oscillators, the results for the simple susceptibility (\ref{chi}) give some qualitative indications. The results illustrated in figure~\ref{fig:zpe} suggest that if all oscillators in the set are brought to a fixed temperature, then the energy removed from the oscillators in reaching their ground states will increase with damping, where the damping level is determined from the measured susceptibilities.

\section{Conclusions}
The quantum damped oscillator can be described using the techniques of macroscopic quantum electrodynamics. Using this approach we have calculated the thermal and zero-point energy of a damped oscillator for a simple model susceptibility. Experimental quantum oscillators will be characterized by susceptibilities that must be measured, just as the electromagnetic susceptibilities of materials must be measured to quantify effects such as Casimir forces. Our analytical results for a model susceptibility show an interesting effect that may also be present in experimental systems. The energy stored in the oscillator, which must be removed to reach the quantum ground state, increases with temperature due to damping of the zero-point energy. By engineering the damping of a set of oscillators, it may be possible to observe this effect in current experimental systems~\cite{oco10,teu11,cha11,asp10,poo12,gro13}.

\section*{Acknowledgements}
We are indebted to J.\ Anders for informing us of the Hamiltonian of mean force. We also thank N.\ Kiesel and M.\ Aspelmeyer for interesting discussions.

\appendix
\section*{Appendix}
\setcounter{section}{1}
Consider a system composed of two interacting parts; the system of interest (\(S\)), and a reservoir (\(R\)).  The total Hamiltonian of this system is of the form, \(\hat{H}=\hat{H}_{S}+\hat{H}_{I}+\hat{H}_{R}\), where \(\hat{H}_{I}\) characterises the coupling (of arbitrary strength) between \(S\) and the reservoir.  We ask the question, \emph{what is the energy of \(S\) in thermal equilibrium}?
\par
A choice of Hamiltonian, \(\hat{H}^{\star}\) that gives the correct equilibrium properties for \(S\) without reference to \(R\) is the Hamiltonian of mean force~\cite{cam09}
\begin{equation}
\hat{H}^{\star}=-\beta^{-1}\log\left(Z_{R}^{-1}\mathrm{Tr}_{R}\left[e^{-\beta\hat{H}}\right]\right)  \label{hmf}
\end{equation}
where \(\beta=1/kT\), and \(Z_{R}=\mathrm{Tr}_{R}[\exp{(-\beta\hat{H}_{R}})]\). The partition function $Z^\star$ associated with $\hat{H}^{\star}$ is then
\begin{equation}   \label{Zstar}
Z^\star=\mathrm{Tr}_{S}\left[e^{-\beta\hat{H}^\star}\right]=\frac{Z}{Z_R},
\end{equation}
where  \(Z=\mathrm{Tr}[\exp{(-\beta\hat{H}})]\) is the total partition function.
It is evident that equilibrium averages of quantities pertaining to \(S\) alone, computed using \(\hat{H}^{\star}\) will be identical to those calculated from the full Hamiltonian \(\hat{H}\).  The factor of \(Z_{R}^{-1}\) within the logarithm plays no role in such a calculation, but is determined by the requirements that (a) when \(\hat{H}_{I}\to0\) then \(\hat{H}^{\star}\to\hat{H}_{S}\); and (b) the free energy  \(F^{\star}\) associated with \(S\) is~\cite{for85}
\begin{equation}
	F^{\star}=-\beta^{-1}\log(Z^{\star})=F-F_{R},  \label{free-energy}
\end{equation}
which is the amount of energy available to do work in a reversible, isothermal change of state of \(S\), including that obtained through decoupling \(S\) and \(R\)~\cite{for85}.  In the case considered in the main text, where the $q$-oscillator plays the role of \(S\), (\ref{free-energy}) will give the correct generalized force (and therefore work done during any isothermal change of state) when \(F^{\star}\) is differentiated with respect to the ``free-oscillation" frequency \(\omega_{0}\), or the quantities \(\gamma_{1,2}\) within the coupling of the oscillator to the reservoir.  Furthermore, when \(\gamma_{1,2}\to0\) then \(F^{\star}\to F_{S}\).
\par
Given the above properties, (\ref{hmf}) is interpreted as the effective Hamiltonian of \(S\) in thermal equilibrium.  In answer to our initial question, the equilibrium average of the associated energy is, using (\ref{Zstar}),
\begin{equation}
	\langle\hat{H}^{\star}\rangle=-\frac{\partial \log(Z/Z_{R})}{\partial\beta}=\langle\hat{H}\rangle-\langle\hat{H}_{R}\rangle,
\end{equation}
where \(\langle\hat{H}_{R}\rangle=-\partial\log(Z_{R})/\partial\beta\) is the equilibrium average of the energy of \(R\) in the absence of any coupling to \(S\).  This is the prescription that was previously used to calculate the Casimir energy density~\cite{phi10} and the thermal energy of a damped harmonic oscillator~\cite{phi12}, the latter of which is given by (\ref{Hq}).


\end{document}